\documentclass[12pt]{article}
\usepackage{amsmath}
\usepackage{graphicx}
\usepackage{mathtools}
\usepackage{mathrsfs}
\usepackage{amsthm}
\usepackage{amsfonts}
\usepackage{amssymb}
\usepackage{enumerate}
\usepackage{natbib}
\usepackage{xcolor}
\usepackage{bm}
\usepackage{lineno}

\usepackage{xr-hyper}
\usepackage{hyperref}
\usepackage{cleveref}

\def\tr{\rm{tr}}

\newcounter{casestudy}
\newcommand{\casestudy}[1]{\refstepcounter{casestudy}\label{#1}}

\newcommand{\blind}{1}
\newcommand{\matern}{Mat\'{e}rn~}

\begin{document}

\def\spacingset#1{\renewcommand{\baselinestretch}%
{#1}\small\normalsize} \spacingset{1}


\if1\blind
{
  \title{\bf Source Reconstruction for Spatio-Temporal Physical Statistical Models}
  \author{Connie Okasaki\thanks{
    This material is based upon work supported by the National Science Foundation Graduate Research Fellowship under Grant No. DGE-1762114}\hspace{.2cm}\\
    Quantitative Ecology and Resource Management Program, U of Washington,\\
    Mevin B. Hooten \\
    Department of Statistics and Data Sciences, The University of Texas at Austin, \\
    and \\
    Andrew M. Berdahl\thanks{AMB was supported by the H. Mason Keeler Endowed Professorship in Sports Fisheries Management.} \\
    School of Aquatic and Fisheries Sciences, U of Washington}
  \maketitle
} \fi

\if0\blind
{
  \bigskip
  \bigskip
  \bigskip
  \begin{center}
    {\LARGE\bf Title}
\end{center}
  \medskip
} \fi

\bigskip
\begin{abstract}
In many applications, a signal is deformed by well-understood dynamics before it can be measured. For example, when a pollutant enters a river, it immediately begins dispersing, flowing, settling, and reacting. If the pollutant enters at a single point, its concentration can be measured before it enters the complex dynamics of the river system. However, in the case of a non-point source pollutant, it is not clear how to efficiently measure its source. One possibility is to record concentration measurements in the river, but this signal is masked by the fluid dynamics of the river. Specifically, concentration is governed by the advection-diffusion-reaction PDE, with an unknown source term. We propose a method to statistically reconstruct a source term from these PDE-deformed measurements. Our method is general and applies to any linear PDE. This method has important applications in the study of environmental DNA and non-point source pollution.
\end{abstract}

\noindent%

\spacingset{1} 
\section{Introduction}
\label{sec:intro}
In many applications, a substance is emitted into a dynamic environment such as a stream or the atmosphere, where it is then passively transported. Measuring concentrations of such a substance is a natural monitoring tool, but inferring emissions from concentrations requires unraveling the dynamics of the transport medium. Often, concentrations can be modeled as the solution to a partial differential equation (PDE), in which case the emissions can be modeled as a source term. 

We present a method using stochastic PDEs (SPDEs) to calculate the conditional distribution of the source term of a linear PDE, given measurements of the PDE solution. These calculations are efficient when the source can be represented as a Gaussian Markov Random Field (GMRF). Our technique builds on the seminal SPDE work of \cite{lindgren2011explicit}, which suggested advection-diffusion modeling as a potential application of their approach. Few studies have extended this application within the SPDE context. The closest in spirit are \cite{sigrist2012dynamic}, which used an equivalent integro-difference equation; and \cite{sigrist2015stochastic}, which used Fourier methods to build a spatio-temporal advection-diffusion model. However, these studies largely demonstrate the case of a single drift vector rather than an entire advection field. Outside the SPDE context, spatial and spatio-temporal statistical models of concentration within dynamic mediums are more common, but often either ignore transport and emissions effects (e.g. \citealt{moraga2017geostatistical}), or incorporate these effects in a non-mechanistic (e.g. \citealt{cameletti2019bayesian}) or pseudo-mechanistic manner (e.g. \citealt{ver2006spatial}). Here we use explicit advection-diffusion mechanisms, within the SPDE context in order to obtain sparse precision matrices, and can incorporate a spatio-temporally varying advection field.

Our method provides a mechanistic kernel for modeling concentrations, as well as a way to infer emissions from those concentrations. The latter is novel in the field of spatial statistics, and statistically more difficult, because transport distorts the original emissions signal. We therefore focus on the latter application. In Section~\ref{sec:background}, we provide background information: a review of the SPDE approach for generating the \matern{}covariance, a review of the finite element method as it is used in this approach, and an overview of the advection-diffusion-reaction equation. In Section~\ref{sec:methods}, we outline the method we used for source reconstruction, as well as the method we used for quantifying the accuracy of our source reconstructions, a method for constructing spatio-temporal versions of our model, the MCMC method we used for sampling, and finally any specifics regarding our three case studies. In Section~\ref{sec:results}, we describe the results of these three case studies.

\subsection{Motivation}
Our focus on inferring emissions is motivated by several continuing and emerging challenges in environmental science. The first and most direct application is the problem of non-point source air and water pollution. There is a long history of investigation into the inverse source problem in these domains (e.g. \citealt{panofsky1969air}), but
study has been largely focused on the more mathematically tractable point-source case (e.g., \citealt{stockie2011mathematics}). Nevertheless, non-point sources are a major concern for regulators, in part because the traditional policies used to regulate pollution are difficult to implement for these sources \citep{xepapadeas2011economics}. 

Another major motivation of this work is environmental DNA\footnote{Environmental DNA is particles of DNA shed from an organism into the environment. These particles can be used to passively detect the presence of particular species \citep{rees2014detection}, or in certain circumstances can be used to estimate biomass \citep{rourke2022environmental}. In theory, provided accurate parameters describing the physical-chemical dynamics of environmental DNA, one could use it to derive a spatio-temporal map of biomass using only non-invasive sampling.}\label{f:eDNA} (eDNA). The eDNA literature has made great strides in the last decade and techniques such as DNA metabarcoding \citep{deiner2017environmental} have seen wide acceptance. More recently, scientists have investigated the use of quantitative polymerase chain reaction analysis (qPCR\footnote{The polymerase chain reaction is a biological reaction used to amplify a particular DNA sequence if it is present in a sample. It is traditionally used as a detector of presence/absence. However, in recent years qPCR has been developed as a technique to estimate not just presence/absence but the \emph{amount} of a particular DNA sequence present in the original sample.}\label{f:qpcr}) to estimate the abundance of organisms (e.g., \citealt{doi2015use},\citealt{doi2017environmental}). Once the dynamic properties of eDNA (shedding rate, decay rate, etc) can be quantified in field settings, eDNA offers a promising tool for non-invasive abundance estimation \citep{harrison2019predicting}. Our method offers a mathematical framework in which many different dynamic properties can be incorporated to infer spatial and spatio-temporal abundance.

The inverse source problem is broad, and includes many problems beyond these two motivating applications. Other literature in this field includes inferring sources of heat \citep{el2002inverse}; acoustic waves \citep{alves2009iterative}; and electromagnetic signals, particularly EEG data \citep{grech2008review}. We do not consider these applications in this paper, but the adaptation of our method to these domains is a promising area of future study. 

\section{Background}
\label{sec:background}

\subsection{Statistical Background}
\label{ssec:background-stats}
We assume that the source, or emissions, $f(\bm{{s}}) \sim GP(\mu(\bm{{s}}),k(\bm{{s}},\bm{{s}}'))$ is a Gaussian process with mean function $\mu$ and covariance function $k$. We assume that $f(\bm{{s}})$ cannot be measured directly, and is the primary target of inference. Furthermore, we assume that $f$ can be represented using a Gaussian Markov Random Field (GMRF; for general reference see \cite{rue2005gaussian}) so that for any particular finite set of points $\bm{{s}}$, $\bm{{f}}(\bm{{s}}) \sim \mathcal{N}(\bm{{\mu}}(\bm{{s}}),\bm{{Q}}(\bm{{s}},\bm{{s}})^{-1})$ with a sparse precision matrix $\bm{{Q}}$. In the special case that $k$ is the Mat\'{e}rn covariance function,
\begin{equation}
k(\bm{{s}},\bm{{s}}') = \frac{\sigma^2}{2^{\nu-1}\Gamma(\nu)}(\kappa||\bm{{s}}'-\bm{{s}}||)^\nu K_\nu(\kappa||\bm{{s}}'-\bm{{s}}||),
\end{equation}
\cite{lindgren2011explicit} provided a method for directly generating a sparse $\bm{{Q}}$ for certain integer and half-integer values of the smoothness parameter $\nu$. Their approach for doing so was based on applying the finite element method (FEM) to a particular set of nested diffusion equations whose solutions are governed by the Mat\'{e}rn covariance function. We use the same method to translate from emissions to concentration, using a different PDE (see Equation~\ref{eq:adv-diff-t}).

\subsection{The Advection-Diffusion-Reaction Equation}
\label{ssec:adv-diff}

We now consider the advection-diffusion-reaction equation in more detail. Suppose that we seek to infer the source of a chemical, but only have measurements of chemical concentration in a fluid medium. The chemical flows with, and disperses through, that medium, and possibly decays linearly over time. The chemical concentration is then described by the equation:
\begin{equation}
    \dfrac{\partial u(\bm{{s}},t)}{\partial t} + \nabla\cdot(\bm{{v}}(\bm{{s}},t)u(\bm{{s}},t)) - \nabla\cdot(D\nabla u(\bm{{s}},t)) + ru(\bm{{s}},t) = f(\bm{{s}},t), \label{eq:adv-diff-t}
\end{equation}
where $\nabla$ is the del operator for the spatial dimensions, $D$ is the diffusion coefficient, $r$ is the decay rate, $\bm{{v}}(\bm{{s}},t)$ is the advection field, assumed here to be known, $f(\bm{{s}},t)$ is the source function, a general Gaussian process in space and time, and $u(\bm{{s}},t)$ is the solution, or concentration function. If we assume that $f(\bm{{s}},t) = f(\bm{{s}})$ is homogeneous in time and that $u(\bm{{s}},t) = u(\bm{{s}})$ has reached steady state, the PDE simplifies to 
\begin{equation}
   \nabla\cdot(\bm{{v}}(\bm{{s}})u(\bm{{s}})) - \nabla\cdot(D\nabla u(\bm{{s}})) + ru(\bm{{s}}) = f(\bm{{s}}).\label{eq:adv-diff-steady}
\end{equation}
For example, this model may be appropriate when trying to estimate the origin of non-point source pollution (e.g., runoff pollution); the source of isotopes found in water samples\footnote{Many elements found in nature have multiple naturally occurring and scientifically useful isotopes, such as tritium ($^3$H), radiocarbon ($^{14}$C), and oxygen-18 ($^{18}$O). Such isotopes can be naturally enriched by environmental processes such as evaporation and can therefore be used to identify the source of water flow (groundwater vs. snow pack vs. rainfall; e.g. \citealt{hao2019stable}) or match a sample with a particular location (e.g. \citealt{rachel2008tracking}).}; or the biomass of fish from eDNA measurements. This inverse problem has been considered in applied math studies, and is usually solved either in the case where the chemical is assumed to originate with unknown strength from a small finite number of unknown point locations \citep{el2005identification,el2007inverse,lushi2010inverse}, or using regularization techniques from the inverse problems literature that do not admit a probabilistic interpretation and do not include an estimate of uncertainty \citep{porter1982holography,engl1996regularization,yan2008method}.

Previous statistical work has examined the advection-diffusion equation in similar contexts using integro-difference equations \citep{sigrist2012dynamic} and Fourier methods \citep{sigrist2015stochastic}. 
\cite{stroud2010ensemble} used the finite difference method to account for these practicalities but used an ensemble Kalman filter method to conduct inference, rather than approaching the problem from an SPDE perspective.

\subsection{The Finite Element Method}
\label{ssec:fem-background}

In general, discretization methods translate partial differential equations into systems of linear equations
\begin{equation}
\bm{{K}}\bm{{u}} = \bm{{L}}\bm{{f}}.\label{eq:FEM}
\end{equation}
In the finite element method, $\bm{{K}}$ is known as the stiffness matrix and $\bm{{L}}$ is known as the mass matrix. In the finite difference method, often $\bm{{L}} = \bm{{I}}$. Essential boundary conditions (i.e., those that are not automatically satisfied; usually Dirichlet) modify this basic equation. We demonstrate this modification in the Appendix. We refer to $\bm{{u}}$ as the solution vector and $\bm{{f}}$ as the source vector, with precision matrices $\bm{{Q}}_u$ and $\bm{{Q}}_f$. 

Our methodology for generating a sparse precision matrix $\bm{{Q}}_u$ given $\bm{{Q}}_f$ is essentially the same as in \cite{lindgren2011explicit}. We generalize their notation and define $\mathcal{L}$ to be any linear partial differential operator and $\mathcal{B}$ to be a general Dirichlet, Neumann, or Robin boundary condition, to produce a general formulation of a PDE
\begin{equation}
\begin{split}
    \mathcal{L}u(\bm{s}) & = f(\bm{s}) \qquad\mbox{for } \bm{{s}}\in \Omega \label{eq:general-pde} \\
    \mathcal{B}u(\bm{s}) & = c(\bm{s}) \qquad\mbox{for } \bm{{s}}\in \partial\Omega.
\end{split}
\end{equation}
Neumann boundary conditions are the natural boundary conditions in our analyses and we use them throughout, but Dirichlet boundary conditions could also be used, as could non-zero Neumann boundary conditions and Robin boundary conditions. Boundary conditions may also be specified in terms of lower-dimensional random fields (see Appendix for details).

The finite element method discretizes Equation~\ref{eq:general-pde} by finding a weak solution using a specified set of test and trial functions. In \cite{lindgren2011explicit}, and in this study, both classes are spanned by a series of piecewise linear ``bump'' functions defined to be 0 over most of the domain, and 1 at one specific node chosen from a set of finite points in space. Other classes may be advantageous for particular SPDEs, including the advection-diffusion equation (which in future work may benefit from the Streamline-Upwind Petrov Galerkin bases; \citealt{brooks1982streamline}) and Maxwell's equations (which benefit from the use of specialized bases such as the N\'{e}d\'{e}lec elements; \citealt{monk2003finite}). For a more comprehensive review of the FEM in general see \cite{brenner2008mathematical}, or for an accessible review of the FEM applied to the Mat\'{e}rn equation see \cite{bakka2018solve}. 

Conceptually, the finite element method serves to translate the PDE in Equation~\ref{eq:general-pde} into a matrix equation such as Equation~\ref{eq:FEM}. Classically, one then solves for $\bm{{u}} = \bm{{K}}^{-1}\bm{{L}}\bm{{f}}$. \cite{lindgren2011explicit} observed that in doing so, we imply that if $\bm{{f}}$ is multivariate normal (MVN) with mean $\bm{{\mu}}_f$ and precision $\bm{{Q}}_f$ then $\bm{{u}}$ is MVN with mean $\bm{{K}}^{-1}\bm{{L}}\bm{{\mu}}_f$ and precision $\bm{{K}}^{T}\bm{{L}}^{-T}\bm{{Q}}_f\bm{{L}}^{-1}\bm{{K}}$. To ensure sparsity of the resulting matrix, we replace $\bm{{L}}$ with a diagonal matrix $\tilde{\bm{{L}}},$ formed by summing the rows of $\bm{{L}}$ and placing the resulting sums along the diagonal. 

\section{Methods}
\label{sec:methods}

\subsection{Finite Element Calculations}
\label{ssec:methods-fem}

Although the finite difference method would produce conceptually identical results, we use the finite element method to compute discretized spatial derivatives. Because the stiffness and mass matrices are standard components of finite element analysis in engineering, they can be calculated by leveraging existing FEM software such as the Finite Element Computational Software (FEniCS \citep{alnaes2015fenics}. We rely on FEM software to implement our stiffness and mass matrices, to improve reproducibility and facilitate generalizability. We assume that the FEM approach is appropriate for any SPDE in which the stochastic source term is well-behaved and in which the FEM equation converges to the true solution in the deterministic case.

\subsection{Source Reconstruction}
\label{ssec:kriging-step}
Our basic observation is that physical meaning can sometimes be ascribed to $f(\bm{{s}})$ and that the finite element representation $\bm{{f}}$ can be calculated by inverting Equation~\ref{eq:FEM} in the reverse direction $\bm{{f}} = \bm{{L}}^{-1}\bm{{K}}\bm{{u}}$. Our method for source reconstruction given a set of parameters (which generally must be estimated) therefore follows essentially the same logic as classical kriging. We assume that the solution $\bm{{u}}$ is observed through some set of linear functionals which can be discretized to a matrix $\bm{{A}}$. We therefore observe $\bm{{y}}=\bm{{Au}}+\bm{{\epsilon}}$ with i.i.d. Gaussian noise. 
%
%
The conditional field $\bm{{u}}|\bm{{y}}$ is therefore
\begin{equation}
    \bm{{u}}|\bm{{y}} \sim N\left(\bm{{\mu}}_u + \frac{1}{\sigma^2}(\bm{{Q}}_u + \frac{1}{\sigma^2_\epsilon}\bm{{A}}^T\bm{{A}})^{-1}\bm{{A}}^T(\bm{{y}} - \bm{{A}}\bm{{\mu}}_u),(\bm{{Q}}_u + \frac{1}{\sigma_\epsilon^2}\bm{{A}}^T\bm{{A}})^{-1}\right).
\end{equation}
This is a standard kriging computation for a GMRF (Eq. 2.15 of \citealt{rue2005gaussian}). However, we propose to take the additional step of kriging $\bm{{f}}|\bm{{y}}$ by observing that $\bm{{f}}|\bm{{y}} = \bm{{L}}^{-1}\bm{{K}}\bm{{u}}|\bm{{y}}$ and therefore
\begin{equation}
    \bm{{f}}|\bm{{y}} \sim N\left(\bm{{L}}^{-1}\bm{{K}}\bm{{\mu}}_{u|y},\bm{{L}}^{-1}\bm{{K}}(\bm{{Q}}_u + \frac{1}{\sigma_\epsilon^2}\bm{{A}}^T\bm{{A}})^{-1}\bm{{K}}^T\bm{{L}}^{-1}\right).
\end{equation}
To facilitate numerical calculation we use the Cholesky factors of $\bm{{Q}}_u$ and $\bm{{Q}}_u + \frac{1}{\sigma_\epsilon^2}\bm{{A}}^T\bm{{A}}$. We calculate these Cholesky factors using \texttt{CHOLMOD} functions \citep{chen2008algorithm} provided by the \texttt{sksparse} library in Python. Specifically, we calculate the Cholesky factor for $\bm{R}^T\bm{R} = \bm{{Q}}_u$, and use the sparse Cholesky update methods provided by \texttt{CHOLMOD} to update that factor to account for the addition of a low-rank matrix $\frac{1}{\sigma^2_\epsilon}\bm{{A}}^T\bm{{A}}$. Note that this update step is computationally difficult when implementing our model in space-time.

Linear regression on $\bm{{f}}$ can be accommodated within this kriging framework; this calculation is shown in the Appendix. 

\subsection{Accuracy Quantification}
\label{ssec:accuracy}

To determine the accuracy of our method for mechanistic kriging, we quantified error using the $L^2$ norm over the domain (excluding any buffer regions) relative to the kriging estimator: 
\begin{equation}
    L^2(x|y) = \sqrt{\int_\Omega (x|y - \mu_{x|y})^2d\Omega}.
\end{equation}
In simulations, we calculated the integral over the finite element basis functions using \mbox{FEniCS} native integration functions. However, assuming the mesh has cells of approximately equal extent, and assuming that all hyperparameters are known perfectly so that all error is produced by Gaussian process variability and observation error, we may approximate the error as
\begin{equation}
    L^2(x|y) \approx \sqrt{\frac{V}{M}\rm{tr}\left(\bm{{I}}_{\rm int}\bm{{\Sigma}}_{\bm{{x}}|\bm{{y}}}\bm{{I}}_{\rm int}\right)},
\end{equation}
where $\bm{{x}}$ represents either $\bm{{f}}$ or $\bm{{u}}$, $V$ is the volume of the interior region $\Omega$, $\bm{{\Sigma}}_{\bm{{x}}|\bm{{y}}}$ is the conditional covariance of vector $\bm{{x}}$ given the vector $\bm{{y}}$ of observations of $\bm{{u}}$, $\bm{{I}}_{\rm int}$ is a diagonal matrix whose entries are the fractions of each basis function contained within $\Omega$, and $M$ is $\tr(\bm{{I}}_{\rm int}).$

Additionally, to quantify the rate of convergence of error, we consider the possibility of observing the same set of of locations $\bm{{A}}^T\bm{{A}}$ a total of $n$ times, resulting in a precision matrix
\[
\bm{{Q}}_{\bm{{u}}} + \frac{n}{\sigma^2_\epsilon}\bm{{A}}^T\bm{{A}}.
\]
We relax the interpretation that we actually observe locations an integral number of times so as to allow a derivative with respect to $n$. We reparameterize (with $m$ the number of rows in $\bm{{A}}$), and set $\zeta = \log(mn)$; we also reparameterize to obtain the log of the error.
Differentiating the log-error with respect to $\zeta$, we can obtain the local polynomial convergence rate. Further, if we do not use a buffer zone, thereby allowing us to observe all nodes, we can also show that as $n\to\infty$, the polynomial convergence rate approaches $-1/2$ for both the source and solution functions. This corresponds to a $1/\sqrt{N}$ type asymptotic convergence for the $L^2$ error. We show that this asymptotic behavior is not observed empirically, either due to the inclusion of a buffer region, or due to slow convergence to the asymptote. Instead, the local polynomial convergence rate calculation is more useful in practice. Full calculations are shown in the Appendix. 

\subsection{Spatio-Temporal Modeling}
\label{ssec:spatio-temporal}

Our implementation of time-dependence relies on a finite element method in space and a finite difference method in time. This results in a sparse vector auto-regressive (VAR) model for the spatial variables \citep{wikle2010general}. 
We discretize the time interval $[t_0,t_1]$ into small units of $\Delta t = (t_1-t_0)/N$. We can then derive a matrix equation from any arbitrary SPDE 
\begin{equation}
    \dfrac{\partial u(\bm{{s}},t)}{\partial t} + \mathcal{L}(t)u(\bm{{s}},t) = f(\bm{{s}},t) \label{eq:space-time}
\end{equation}
(where $\mathcal{L}(t)$ is an arbitrary partial differential operator; for example, in Equation~\ref{eq:adv-diff-t}, we have $\mathcal{L}(t)u = \nabla\cdot(\bm{{v}} u)-\nabla\cdot(D\nabla u) + r u$) by first applying a backward Euler discretization in time
\begin{equation}
\begin{split}
    \dfrac{u(\bm{{s}},t+\Delta t) - u(\bm{{s}},t)}{\Delta t} + \mathcal{L}u(\bm{{s}},t+\Delta t) & \approx f(\bm{{s}},t+\Delta t ), \\
    u(\bm{{s}},t+\Delta t) + \Delta t\mathcal{L}u(\bm{{s}},t+\Delta t) & \approx u(\bm{{s}},t) + \Delta tf(\bm{{s}},t+\Delta t),
\end{split}
\end{equation}
and then applying an FEM discretization in space
\begin{equation}
\begin{split}
    \bm{{L}}\bm{{u}}_{t+\Delta t} + \Delta t\bm{{K}}_{t+\Delta t}\bm{{u}}_{t+\Delta t} & \approx \bm{{L}}\bm{{u}}_t + \Delta t\bm{{L}}\bm{{f}}_{t+\Delta t}, \\
    (\bm{{I}} + \Delta t\bm{{L}}^{-1}\bm{{K}}_{t+\Delta t})\bm{{u}}_{t+\Delta t} & \approx \bm{{u}}_{t} + \Delta t\bm{{f}}_{t+\Delta t}, \\
    u(\bm{{s}},t) & \approx (\bm{{I}} + \Delta t\bm{{L}}^{-1}\bm{{K}}_{t+\Delta t})^{-1}(\bm{{u}}_t+\Delta t \bm{{f}}_{t+\Delta t}),
\end{split}
\end{equation}
where all functions of space and time are replaced by vectors in space, subscripted by time, and the operator $\mathcal{L}(t+\Delta t)$ is discretized into $\bm{{K}}_{t+\Delta t}$.
This equation can be solved inductively by
\begin{equation}
\begin{bmatrix} \bm{{u}}_1 \\ \bm{{u}}_2 \\ \vdots \\ \bm{{u}}_N\end{bmatrix}
= 
\begin{bmatrix} \bm{{M}}_1 \\ \bm{{M}}_2\bm{{M}}_1 \\ \vdots \\ \bm{{M}}_N...\bm{{M}}_1\end{bmatrix}\bm{{u}}_0 
+ 
\Delta t
\left[
    \begin{array}{cccc}
    \bm{{M}}_1      &         &        &             \\
    \bm{{M}}_2\bm{{M}}_1      & \bm{{M}}_2       &        &            \\ 
    \vdots & \vdots   & \ddots     &  \\ 
    \bm{{M}}_N...\bm{{M}}_1    & \bm{{M}}_N...\bm{{M}}_2 & \cdots & \bm{{M}}_{N} \\ 
  \end{array}\right]
\begin{bmatrix} \bm{{f}}_1 \\ \bm{{f}}_2 \\ \vdots \\ \bm{{f}}_{N}\end{bmatrix},
\end{equation}
where $\bm{{M}}_i = (\bm{{I}}+\Delta t\bm{{L}}^{-1}\bm{{K}}_{t_0+i\Delta t})^{-1}$.
The above Toeplitz matrix has sparse inverse
\begin{equation}
\bm{{R}} = \frac{1}{\Delta t}
\left[
    \begin{array}{cccc}
    \bm{{M}}_1       &        &         &  \\
    -\bm{{I}}  & \bm{{M}}_2          &         &  \\ 
         & \ddots & \ddots  &  \\ 
         &        & -\bm{{I}}     & \bm{{M}}_N \\ 
  \end{array}\right].
\end{equation}
A forward Euler discretization yields a very similar matrix, but requires that the time step $\Delta t$ be small enough to ensure numerical stability.

Suppose now that the full, discretized, spatio-temporal process $\bm{{f}}$ has sparse precision matrix $\bm{{Q}}_f$ and that we have a deterministic initial condition $\bm{{u}}_0 = \bm{{0}}$ (accounted for by including a temporal buffer region before the region of interest). Then the precision matrix for $\bm{{u}}$ is
$\bm{{Q}}_u = \bm{{R}}^T\bm{{Q}}_f\bm{{R}}$.
Our initial condition requires us to extend the temporal discretization prior to the first time-of-interest, to avoid boundary effects. Alternatively, the steady-state equation could be used to derive an approximate spatial prior for the initial condition, although this should still be coupled with a (perhaps smaller) buffer region. 

This model uses sparse precision matrices for efficient computation. The block bi-diagonal Cholesky factors allow rapid computation in a high dimensional space, in part because the blocks are themselves sparse matrices. This makes computing likelihoods efficient, scaling linearly in the number of time-steps modeled. The computational difficulty of this model arises from the Cholesky update step (i.e., the factorization of $\bm{Q}_{\bm{u}} + \frac{1}{\sigma^2_\epsilon}\bm{A}^T\bm{A}$), necessary both for inferring the distribution for the emissions or concentration maps, conditional on the observations (the kriging step) and for calculating the marginal likelihood function (the marginalization step). These conditional updates break the structure of our Cholesky factors, resulting in much greater computational burden. In essence, the problem is that information from each observation is transported over time, resulting in long-range dependencies. In densely sampled systems, this problem might be solved by calculating the temporal range of the covariance function, and truncating dependencies after a fixed number of time-steps, thus limiting the bandwidth of the resulting updated Cholesky factor.

\subsubsection{Nested Diffusion Definition of Spatio-temporal \matern Source}
\label{ssec:space-time-matern}
The need for a non-mechanistic spatio-temporal source distribution compatible with our method motivates the following definition of a three-parameter spatio-temporal Mat\'{e}rn field
\begin{equation}
\left(\tau \dfrac{\partial}{\partial t} + \kappa^2 - \Delta\right)^{\alpha/2}u(\bm{{s}},t) = \mathcal{W}(\bm{{s}},t).\label{eq:nested-diffusion}
\end{equation}
This is a nested spatio-temporal diffusion equation, where the more general source term $f(\bm{{s}},t)$ is specified to be constant-in-time spatial white noise (denoted $\mathcal{W}(x)$ for consistency with \citealt{lindgren2011explicit}) then the spatial \matern SPDE is the steady state solution. Although the \matern model does not necessarily require interpretation, the interpretation of this nested diffusion model is more straightforward than an alternative model for which the spatial \matern SPDE is also the steady state solution
\begin{equation}
\tau \dfrac{\partial u(\bm{{s}},t)}{\partial t} + \left(\kappa^2 - \Delta \right)^{\alpha/2}u(\bm{{s}},t) = \mathcal{W}(\bm{{s}},t).\label{eq:unnested-diffusion}
\end{equation}
The two equations are equivalent when $\alpha = 2$; both can be seen as a diffusion-decay equation with natural time scale $\tau$. In this circumstance, ``randomness'' in the form of white noise is spread out across space by means of a familiar mechanism. At higher values of $\alpha$ the equations and their interpretations diverge. Equation~\ref{eq:nested-diffusion} can be seen to be a nested form of the diffusion-decay equation; converting from an Eulerian to a Lagrangian perspective, one can imagine random particles entering a system according to $\mathcal{W}(\bm{{s}},t)$ and diffusing, only to continue emitting secondary particles that themselves diffuse, and so on. In contrast Equation~\ref{eq:unnested-diffusion}, even for integer values of $\alpha/2$, involves at minimum fourth-order-in-space operators coupled to a first-order-in-time derivative. Fourth-order equations, while not at all uncommon in specialized mathematical fields, are relatively exotic. This definition aligns with that proposed in \citep{bakka2020diffusion}, with $\alpha_s=2$ and $\mathcal{E}_Q$ white in both space and time.

\begin{figure}
\centering
\includegraphics[width=5.5cm]{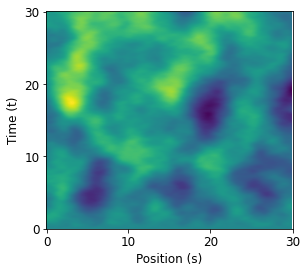}
\caption{The results of a numerical simulation of the proposed spatio-temporal Mat\'{e}rn distribution. Note that this function is both the solution to an SPDE in one context ($u(s,t)$ in Equation~\ref{eq:nested-diffusion}) but that we propose to use it as a source function in another context ($f(s,t)$ in Equation~\ref{eq:space-time}). Parameters used were: $\tau = 2,\alpha = 4, \kappa = 1$, corresponding to $\rho\approx 7$. A spatial buffer zone of length 15 on each side was excluded from this figure. No temporal buffer zone was removed. Space-time was discretized using 751 spatial nodes, and $10^4$ time steps.\label{fig:matern}}
\end{figure}

If we wish to avoid fractional diffusion we must assume $\alpha = 2n$ for $n\in \mathbb{Z}_{+}$, for which we obtain 
\begin{equation}
\bm{{Q}}_u^{(2n)} = \frac{\bm{{R}}^T}{\Delta t}^{n}\left(\bm{{L}}\otimes \bm{{I}}\right)\frac{\bm{{R}}}{\Delta t}.
\end{equation}
This results in $\bm{{Q}}_f = \bm{{L}}\otimes \bm{{I}},$ which is straightforward to calculate. For example, Figure \ref{fig:matern} shows the results of a numerical simulation of this distribution. Although a spatial buffer-zone was included in these calculations and removed from the plot, the temporal buffer-zone was not removed, to demonstrate the transient behavior imposed by our deterministic initial condition. A small zone with length on the order $\tau$ displays lower spatio-temporal variance than the remainder of the plot, but no differences appear to persist beyond that. For this set of parameters, we used $\Delta t = 0.05$ and $2000$ time steps. With 751 spatial nodes, simulation required $\approx 21$ seconds on a desktop computer with a 3.2 GHz 6-core processor and 32 GB of RAM.

\subsection{Parameter Estimation}
\label{ssec:parameter-est}

To estimate the parameters of our air pollution model (see Section~\ref{ssec:pm25}) we used MCMC. To facilitate a Gibbs update for the variance, we specified inverse-gamma priors for $\sigma^2_f$ the variance of the emissions function, and we specified gamma priors for $\rho,v,D,$ and $r$ (respectively, the range; ratio of observation variance $\sigma^2_\epsilon$ to emissions variance $\sigma^2_f$; diffusion coefficient; and decay coefficient). All other parameters are updated in Metropolis-Hastings steps.

We marginalized over the concentration and emission vectors $\bm{{u}}$ and $\bm{{f}}$, but used a second-stage sampler to draw from the full-conditional distributions of $\bm{{u}}$ and $\bm{{f}}$ to compute their posterior means and pointwise standard deviations. 

To treat the range as a parameter we used the empirical equation  $\rho=\sqrt{(8\nu)}/\kappa$ which relates range to the \matern parameters \citep{lindgren2011explicit} and transformed the log-likelihoods accordingly. To use Gibbs sampling for $\sigma^2_f$ we reparameterized to the variance ratio: $v = \frac{\sigma^2_\epsilon}{\sigma^2_f}$. We transformed the log-likelihoods accordingly. Due to the identifiability problems with the \matern $\alpha$ parameter \citep{lindgren2011explicit}, we assumed \textit{a priori} that $\alpha=2$ and therefore $\nu = 1$, and that the prior mean of $\bm{f}$ was everywhere zero. 

To verify the performance of this MCMC algorithm, we also conducted MCMC in our 1-D non-temporal model (see Section~\ref{ssec:adv-diff-regression}). In this model we also reparameterized the hyperparameter $\sigma^2_\beta = \sigma^2_fv_\beta$ to allow a Gibbs update for the emissions variance. Our air pollution model did not include covariates and therefore did not require this parameter. We also used different numbers of iterations in the two examples. Otherwise the algorithms were identical. Note that we did not conduct MCMC in our 1-D spatio-temporal model (see Section~\ref{ssec:space-time-adv-diff}) due to computational constraints. In this case study, we only demonstrate the mechanics of the kriging step. 

\subsection{Case Studies} 
\casestudy{cs:1D}
\subsubsection{Case Study~\ref{cs:1D}: Inferring a 1-Dimensional Spatial Source}
\label{ssec:adv-diff-regression}

We conducted an analysis in one-dimension of space, for a system at equilibrium, with a source governed by a linear model. 

Suppose we make measurements of runoff pollution $u(s)$ at various points in a 1-D stream. The land alongside the stream may be used for several different purposes; for example, it may be agricultural, residential, or industrial. The primary source of variation in pollutant source $f(s)$ may be due to variations in land use (represented by a factor variable), so that $f(s) = \bm{{x}}(s)^T\bm{{\beta}} + \epsilon(s)$ where $\epsilon$ can be represented as a mean-zero GMRF and $\bm{{\beta}}$ has a normal prior. In this case study, to demonstrate the kriging step, we assumed all dynamic parameters were known and just calculated the kriging step. We applied the finite element method to Equation~\ref{eq:adv-diff-steady}  to obtain $\bm{{K}}$ and $\bm{{L}}$. Then, we calculated the joint conditional distribution for $\bm{{f}}$ and $\bm{{\beta}}$ given measurements of $\bm{{y}} = \bm{{Au}}+\bm{{\epsilon}}$ as described in Section~\ref{ssec:kriging-step} by first calculating the joint conditional distribution of $\bm{{u}}$ and $\bm{{\beta}}$ then inverting Equation~\ref{eq:FEM} to obtain the conditional distribution for $\bm{{f}}$. 

To demonstrate the ``kriging step'' we first made predictions assuming the dynamics were known. Then, to demonstrate the accuracy of this kriging calculation, we calculated the average $L^2$ error across 30 simulations for each of 30 sampling densities ranging from a single data point up to ~200,000 data points. For each sampling density, samples were evenly distributed across the interior of the fixed domain, but not into the buffer region. For the purposes of approximating the error we assumed that we observed nodes uniformly $\bm{{A}}^T\bm{{A}} = \bm{{I}}_{\rm int}$  a fractional number of times equal to $n = N/m$, the true sample size divided by the number of rows of $\bm{{A}}$. Note that the relative error that we estimate here is purely interpolation error -- all parameters are assumed known. 

Finally, we conducted an MCMC analysis (described in Section~\ref{ssec:parameter-est}) to demonstrate the efficacy of our inference in the air pollution case study. In this case study, we ran 4 chains for 5,000 iterations each, with a thinning rate of 5 for a total of 4,000 samples. We also used 5,000 iterations of burn-in. To match our air pollution case study, we specified weakly informative priors. Since we marginalized out the source, solution, and regression coefficients, we calculated these posterior distributions as generated quantities after completing the main MCMC sampling.

The true parameter values used for simulation in this case study can be found in Table~\ref{tab:cs-pars}.

\begin{table}
    \centering
    \begin{tabular}{c|ccccccc}
        Parameter & $\alpha$ & $\rho$ & $D$ & $r$ & $\sigma^2_f$ & $\sigma^2_\epsilon$ & $\sigma^2_\beta$\\ \hline
        Case Study~\ref{cs:1D} & 2 & 2 & 0.75 & 0.2 & 10 & 10 & 25 \\
        Case Study~\ref{cs:space-time} & 4 & 5.3 & 0.25 & 0.05 & 10 & 10 & NA
    \end{tabular}
    \caption{The true parameters used for simulation in Case Studies~\ref{cs:1D} and \ref{cs:space-time}. }
    \label{tab:cs-pars}
\end{table}

\casestudy{cs:space-time}
\subsubsection{Case Study~\ref{cs:space-time}: Inferring a 1-Dimensional Spatio-Temporal Source}
\label{ssec:space-time}
In the previous section, we assumed a PDE in steady-state. Steady-state PDEs assume that dynamics and emissions are constant, or change on a time scale much longer than the time scale over which the steady-state solution is reached. However, in many applications these assumptions do not hold. 
In these cases, a full spatio-temporal model may be necessary. We provide a demonstration of this model, but we note that spatio-temporal models are computationally much more challenging, and we defer a more intensive spatio-temporal application for future work. In particular we demonstrate only the kriging step that is the foundation of these calculations, and not a more intensive MCMC analysis. 

In this case study, we simulate a 1-D domain, with small enough dimensionality that computation remains feasible (101 spatial nodes over 2000 time steps, with simulation taking $\approx$1 second on a desktop computer with a 3.4 GHz 6-core processor and 32 GB of RAM. Full spatio-temporal emissions estimation in higher-dimensional spaces is a promising area of future study. 

The true parameter values used for simulation in this case study can be found in Table~\ref{tab:cs-pars}.

\casestudy{cs:air-pollution}
\subsubsection{Case Study~\ref{cs:air-pollution}: Inferring Sources of Air Pollution in the U.S.}
\label{ssec:pm25}

Finally, we implemented the steady-state advection-diffusion-reaction equation to conduct source inference on PM$_{2.5}$ over the United States for January 1st 2019. We obtained PM$_{2.5}$ data from the EPA \citep{epadata} and parameterized our wind-speed field using NOMADS \citep{rutledge2006nomads}. We assumed a homogeneous prior mean for $\bm{{f}}$, although population density, automobile density, land use, and many other covariates could also be included. We included a buffer region of approximately 4 degrees on all sides, and projected locations from degrees latitude/longitude to kilometers, preserving the distance between longitudes at all latitudes. We also projected advection velocities to ensure that the travel time between two points was preserved, although distances between two points and advection speeds were in general not preserved.

We assumed that the emissions (source function; $f(\bm{{s}})$) were governed by a \matern Gaussian process and that the variance and range of this process were unknown, along with the diffusion and decay coefficients. We fit this model using an MCMC algorithm described in~Section~\ref{ssec:parameter-est}. In this case study we ran 4 chains for 25,000 iterations each, with a thinning rate of 5 for a total of 20,000 samples. We also used 25,000 iterations of burn-in.

We specified weakly informative priors based on existing literature ($\rho$ \citep{song2018using}; $D$ \citep{byun2006review}; $r$ \citep{watson2000fugitive,cao2013evolution}; $\sigma^2_f$ \citep{tucker2000overview}; $v$ \citep{peters2001field}) and understanding of the mechanisms involved.
Prior parameters and quantiles, and posterior statistics can be found in Table~\ref{tab:mcmc}.

\section{Results}
\label{sec:results}

\subsection{1-Dimensional Spatial Source Results}
\begin{figure}
\centering
\includegraphics[width=15cm]{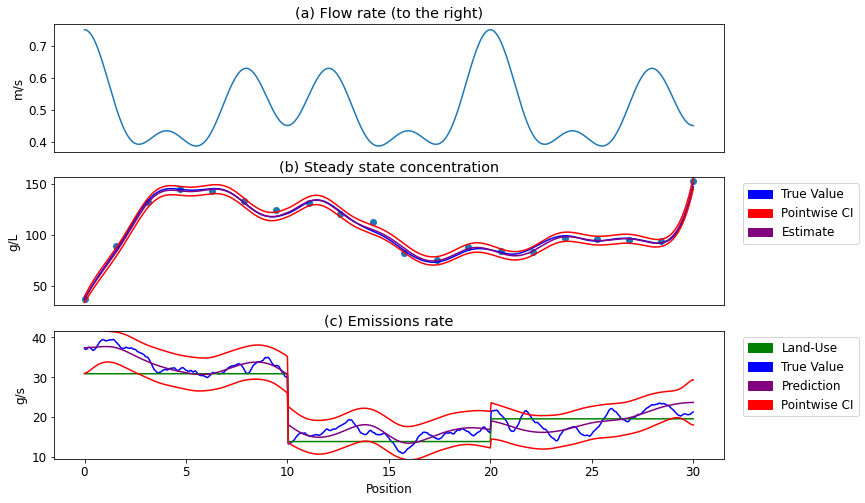}
\caption{The results of a numerical experiment for conducting source reconstruction on the advection-diffusion-reaction equation, with a linear regression on spatial factors. Panel (a) shows the spatially-variable flow rate (always to the right). Panel (b) shows inference on $\bm{{u}}$. Panel (c) shows inference on $\bm{{f}} = \bm{{X}}\bm{{\beta}} + \bm{{\epsilon}}$. Note the discontinuities in $\bm{{f}}$ generated by discontinuities in $\bm{{X}}$. Dots show positions of measurements and their observed values.\label{fig:landuse}}
\end{figure}

Figure \ref{fig:landuse} shows the results of our kriging analysis, with three land-use zones applying differing constant contributions to the source. 
This demonstrates the ability to incorporate a linear regression model into our method and to obtain analytical conditional distributions when dynamics are perfectly known. 

When quantifying the error incurred by kriging in this case study, we found that the local polynomial convergence rate approximations for the $L^2$ error agreed with our empirical calculations at all but the lowest sample sizes ($\lesssim$ 10). We also found that across a wide range of moderate sample sizes, the error converges at a rate slower than $1/\sqrt{N}$. Specifically, we observed a convergence rate of approximately $N^{-0.4}$ when estimating concentrations (i.e., PDE solutions), and an error convergence rate of approximately $N^{-0.1}$ when estimating emissions (i.e., PDE sources). 
Hypothesizing that lower rates of convergence are the result of information being obscured by the dynamics, particular by diffusion and decay, and of observation error being too large to effectively ``see past'' these obfuscation, we performed the same set of simulations with the diffusion rate ($D$), decay rate ($\lambda$), and observation variance ($\sigma^2_\epsilon$) all reduced by a factor of 10. In this case, we observed faster convergence rates of $N^{-0.41}$ and $N^{-0.19}$ respectively. 

\begin{figure}
\centering
\begin{tabular}{cc}
\includegraphics[width=7.5cm]{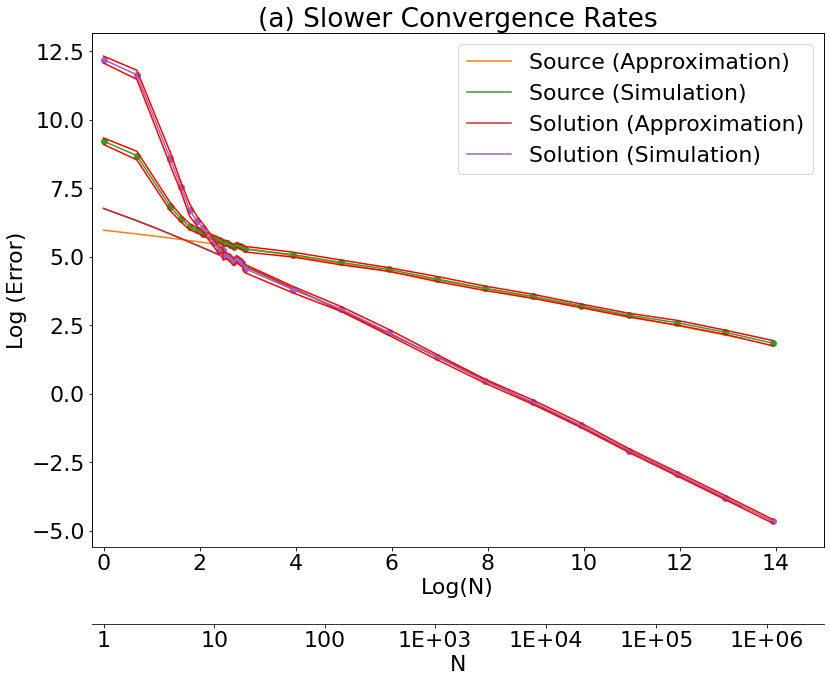} &
\includegraphics[width=7.5cm]{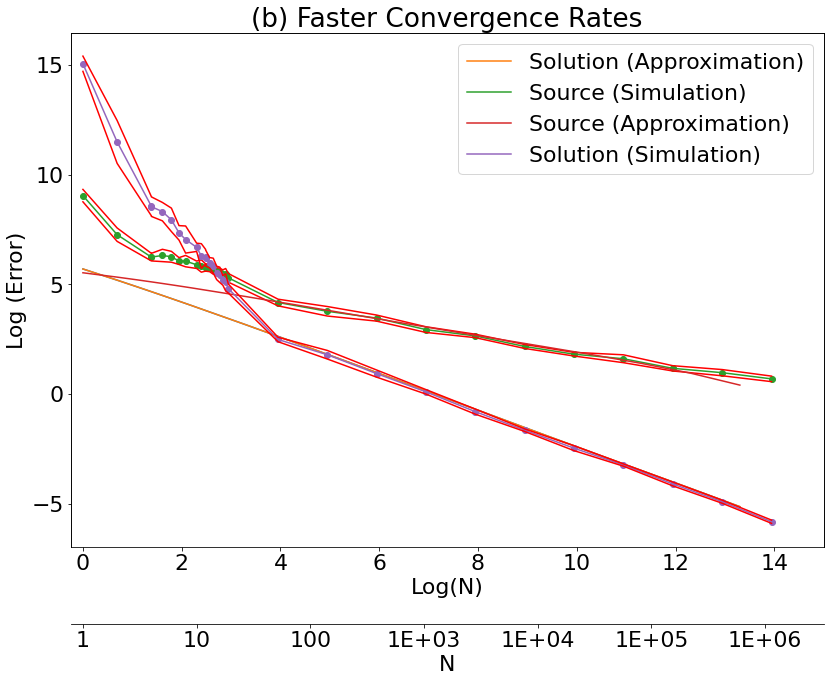}
\end{tabular}
\caption{\label{fig:accuracy}Empirical vs. approximate plots showing the convergence of error for our method when estimating both the source and solution of a PDE from solution-only measurements in Case Study~\ref{cs:1D}. Note that our analytical approximations underestimate the error initially, but become accurate at a sample size of $N < 10$. In the first simulation (a), $D = 0.75, r = 0.2,$ and $\sigma^2_{\epsilon} = 5$, while in the second (b) $D = 0.075, r = 0.02,$ and  $\sigma^2_{\epsilon} = 0.5$.}
\end{figure}

\begin{figure}
\centering
\includegraphics[width=10.5cm]{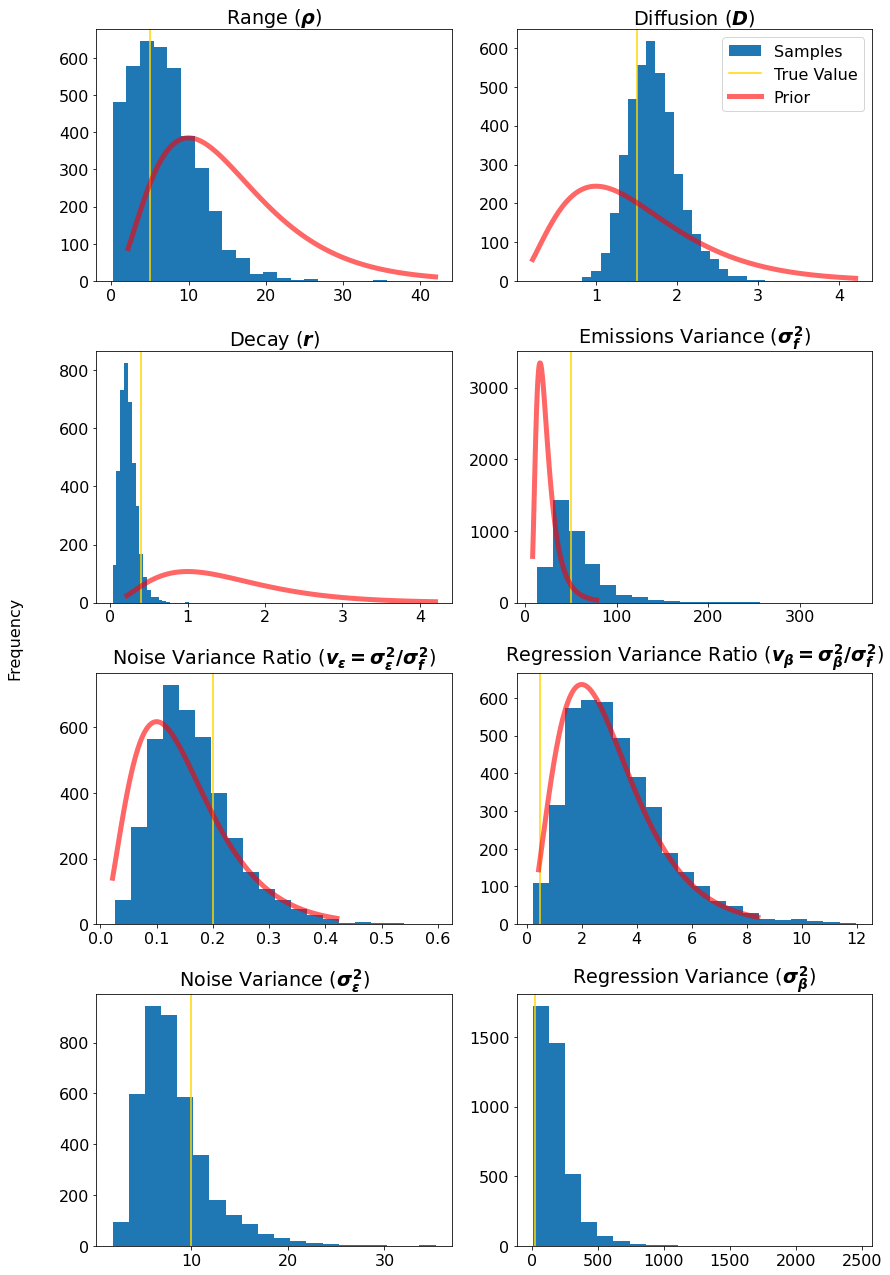}
\caption{The results of Case Study~\ref{cs:1D}'s MCMC analysis. Prior PDF had been scaled up by the number of samples to be comparable to the histogram. \label{fig:mcmc-example}}
\end{figure}

\begin{figure}
\centering
\includegraphics[width=15cm]{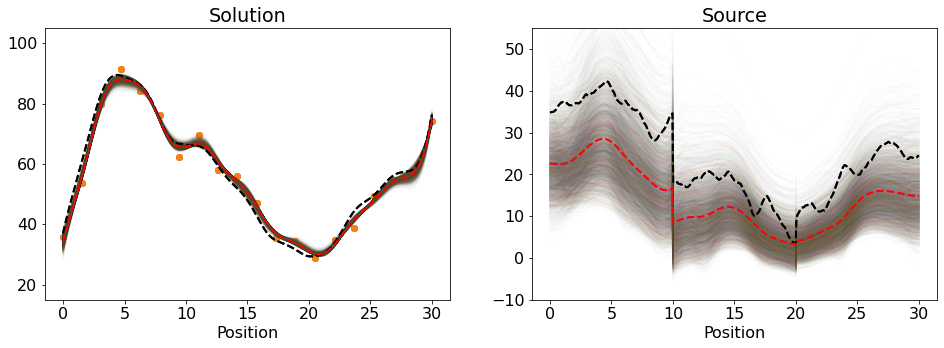}
\caption{The results of Case Study~\ref{cs:1D}'s MCMC analysis. Black dotted lines represent the true simulated source and solution functions. Red dotted lines represent the pointwise posterior mean. Background lines represent the kriging estimate from each MCMC sample. Note that the solution (or concentration) is estimated precisely but the source (or emissions) is more uncertain. Both functions lie largely within the range of variation among samples.  \label{fig:mcmc-example-pred}}
\end{figure}

\begin{figure}
\centering
\includegraphics[width=16cm]{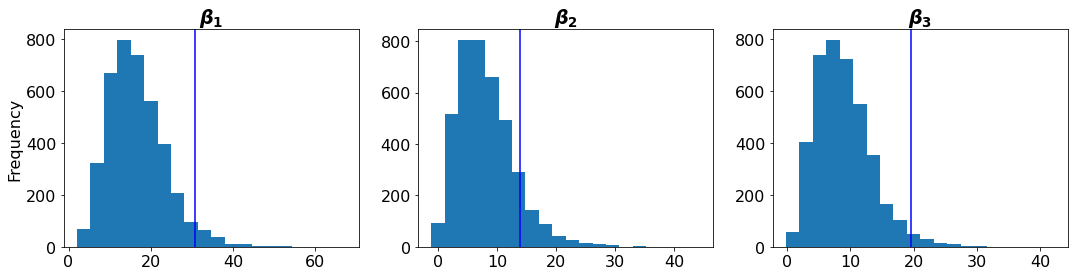}
\caption{The generated distributions for the source, solution, and regression coefficients given the posterior samples of the parameters. The vertical line represents the true value of the coefficient. Note that all three coefficients are underestimated on average, perhaps due to oversmoothing in the kriging step. \label{fig:mcmc-example-coefficients}}
\end{figure}

Upon conducting MCMC we found that some parameters could be inferred well while others suffered from issues of identifiability. The posteriors for the diffusion, decay, range, and emissions variance (shown in Figure~\ref{fig:mcmc-example}) improved noticeably over the priors, although the decay showed a slight tendency toward underestimation. Range, diffusion, decay, and emissions variance were fairly well identified, while the two variance ratios were poorly identified. The unidentifiability of the regression variance ratio had little effect on the regression coefficients however. These three regression coefficients (shown in Figure~\ref{fig:mcmc-example-coefficients}) were somewhat underestimated, likely due to oversmoothing in the kriging estimates. 

Due to the underestimation issues in the regression coefficients, we recommend interpreting model output largely in terms of overall trends and predictions of the source/solution functions. For example, in Figure~\ref{fig:mcmc-example-pred}, the model successfully identifies that the left-most zone has generally higher emissions than two zones on the right. In cases where accurate estimation of regression coefficients is a high priority, we recommend specifying non-Gaussian priors such as the Laplace or Horseshoe priors, and updating the coefficients in the MCMC step. Ultimately, the model allows us to infer the solution with high accuracy, and although the source is more uncertain, the inference remains accurate, with the true source falling well within the range of variation among the samples.

\subsection{1-Dimensional Spatio-Temporal Source Results}

\label{ssec:space-time-adv-diff}
\begin{figure}
\centering
\includegraphics[width=16cm]{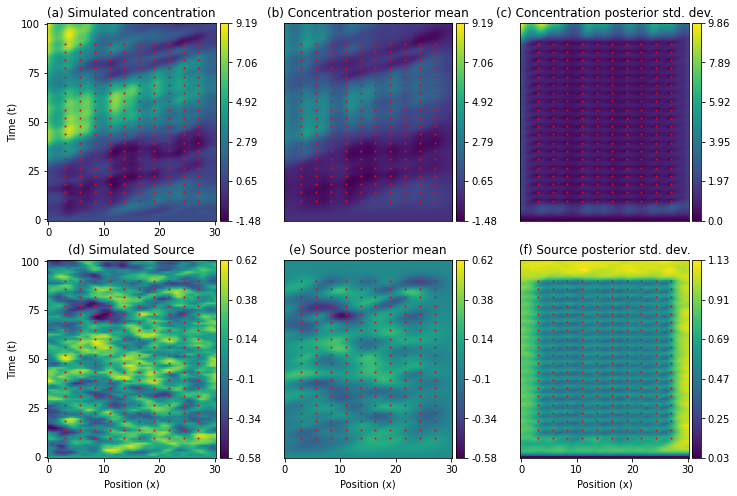}
\caption{The results of Case Study~\ref{cs:space-time}. The source process was spatio-temporal Mat\'{e}rn as defined in Section~\ref{ssec:space-time-matern} with parameters: $\tau = 2,\alpha = 4, \kappa = 1$. Flow rate was constant in time, and varied in space as shown in figure \ref{fig:landuse}. Red points denote sampling locations. A spatial buffer zone of length 5 on each side is not included in the plot. No temporal buffer zone was removed. \label{fig:adv-diff-time}}
\end{figure}

Figure~\ref{fig:adv-diff-time} shows the results of our analysis when we add a temporal dimension to the 1-D system analyzed in Numerical Case Study~\ref{cs:1D}. We simulated the advection-diffusion equation with a spatio-temporal Mat\'{e}rn source as defined in equation (\ref{eq:nested-diffusion}). We simulated 10 sensors equidistant along the $s$-axis, and simulated 20 measurements each, equidistant in time. 

The solution (concentration) estimate is more accurate than the source (emissions) estimate, because this is the scale at which we are able to take direct measurements. The source estimate displays a greater degree of smoothing due to the obscuring effects of diffusion and advection. Because the material detected in a concentration measurement could have originated from any point nearby or upstream, a larger number of measurements are required to provide equivalent resolution on the source function. As more measurements are recorded, the reconstruction converges to the simulated source. Our method can be thought of as a filter that treats concentration measurements as an integral over previous, mostly upstream sources. That is, a measurement of concentration at time $t$ only provides information about the source process up to time $t$, and provides much stronger information from upstream locations (due primarily to advection) than from downstream locations (due to diffusion). Information about future or downstream source values is primarily obtained from the correlation structure of the source process. Figure \ref{fig:adv-diff-time} panel (f) demonstrates this behavior on its margins, where downstream and future uncertainties are higher than upstream and past uncertainties. 

\subsection{Air Pollution Results}

\begin{figure}
\centering
\includegraphics[width=16cm]{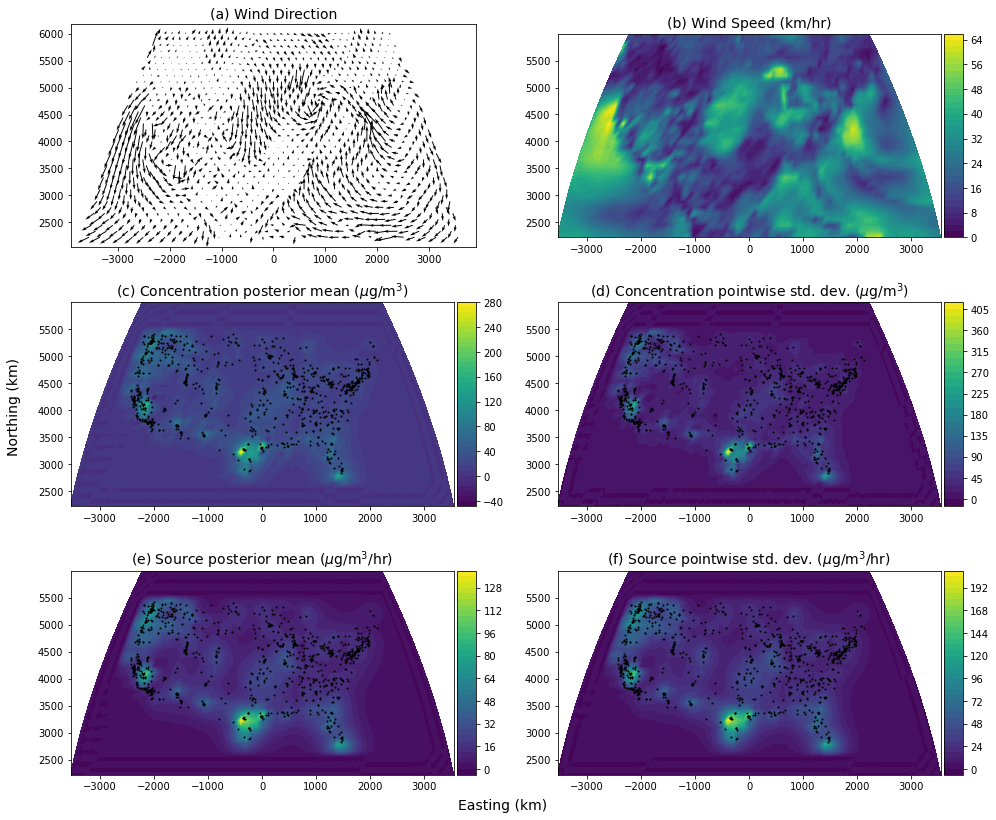}
\caption{The main results of Case Study~\ref{cs:air-pollution}, parameterized with the NOMADs climate model \citep{rutledge2006nomads}, and using EPA PM 2.5 measurements \citep{epadata}. Advection field (a; upper left); advection speed (b; upper right); PM 2.5 map (c; middle left); PM 2.5 map posterior standard deviation (d; middle right); PM 2.5 emissions map (e; lower left); PM 2.5 source map posterior standard deviation (f; lower right). The buffer region (approximately 4 degrees wide) is set to zero to avoid showing boundary anomalies. Dots represent locations of data points. \label{fig:us}}
\end{figure}

The results of this analysis can be visualized in Figure~\ref{fig:us} and Table~\ref{tab:mcmc}. Uncertainty was quantified using the square root of the diagonal elements of the posterior covariance matrix (i.e., the pointwise posterior standard deviations). Posterior uncertainty was lower around the main landmass of the continental US, because this is where the data were collected, and regions of high estimated emissions and concentration were clustered around some regions of high population including Texas, California, Florida, and the Pacific Northwest. Some anomalies are present in the results, notably some rare negative
predictions for PM 2.5 concentrations. These are non-physical, but occur almost entirely outside the sampling region, likely due to the artificial oscillatory behavior induced by the FEM discretization of advection. These could possibly be reduced or removed by using Petrov-Galerkin elements or a full spatio-temporal model, but we leave these possibilities for future work.

\begin{table}
    \centering
    \begin{tabular}{c|c|c|c|c}
        Parameter & Post. Mean & Post. Std. Dev. & Prior & Prior Quantiles (.025,.975) \\ \hline
        $\rho$    & 456 km  & 38 km & $\alpha=4,\beta=0.1$ & (10,80) \\
        $D$ & 8.2km$^2$/hr & 2.8km$^2$/hr &$\alpha=8.5,\beta=1$ & (3.6,14.4)\\
        $r$ & 0.77/hr& 0.42/hr &$\alpha=1.36,\beta=2.94$ & (0.027,1.5)\\
        $\sigma_f^2$  & 32 $(\mu$g/m$^3)^2$ & 44 $(\mu$g/m$^3)^2$ & $\alpha=1.1,\beta=3.9$ & (1,100) \\
        $v$ & 2.1  & 1.7  &$\alpha=1.1,\beta=0.13$ & (0.33,30) \\
        $\sigma_\epsilon^2$ & 29.6 $(\mu$g/m$^3)^2$ & 1.7 $(\mu$g/m$^3)^2$ & NA & NA \\
        $\sigma_f$ & 5.0 $\mu$g/m$^3$ & 2.7 $\mu$g/m$^3$ & NA & NA \\
        $\sigma_\epsilon$ & 5.4 $\mu$g/m$^3$ & 0.16 $\mu$g/m$^3$ & NA & NA \\
    \end{tabular}
    \caption{Posterior mean and standard deviations for each parameter of the advection-diffusion-reaction source model of U.S. air pollution. Also the priors, and prior quantiles for those parameters that had them. The bottom three parameters are generated quantities.}
    \label{tab:mcmc}
\end{table}

The units of emissions in Case Study~\ref{cs:air-pollution} are $\mu$g/m$^3$/hour, because concentration is measured as a density, in units of $\mu$g/m$^3$. Emissions measured in $\mu$g/hour would be more intuitive, but would require making a homogeneity assumption to account for the vertical distribution of PM$_{2.5}$ (we use this approach to approximate a prior distribution for $\sigma^2_f$), or directly modeling vertical distribution with a fully 3-D model. Such a model is not difficult to mesh using FEniCS but is much larger and more computationally intensive, and we leave this extension for future work. 

\section{Conclusion}
\label{sec:conclusion}

We presented a method for SPDE-based inference on the source of a linear PDE. We demonstrated the use of this method for reconstructing the source of an advection-diffusion-reaction process in a 1-D non-temporal setting, a 1-D spatio-temporal setting, and in a 2-D non-temporal case study of air pollution in the United States. We found that the method worked well to estimate concentrations, exhibiting close to a standard $1/\sqrt{N}$ convergence rate. We also found that the method worked well to estimate emissions, but with a reduced error rate that we hypothesized to be due to the destruction of information by diffusion processes. When applied in a case study of realistic complexity, we found that reasonable predictions were made, but we occasionally observed anomalies such as negative concentration. Although these were limited in spatial extent and had a negligible impact on the overall analysis, we hypothesize that these artifacts could be reduced or eliminated in future work by using Petrov-Galerkin elements, or a full spatio-temporal model. We provide a framework through which our model can be extended to space-time, and through which efficient likelihood calculations can be conducted in space-time, but due to the challenges of conducting the marginalization step in space-time (specifically, maintaining sparsity when updating the Cholesky factor of $\bm{Q}_{\bm{u}}$ to add in $\frac{n}{\sigma^2_\epsilon}\bm{A}^T\bm{A}$), leave further spatio-temporal modeling to future work. 

Possible applications of our method include inferring the source of environmental DNA measurements, non-point source pollution, or stable isotopes. Other important linear PDEs include Maxwell's equations and the wave equation. Using these equations, our method could be applied to MRI measurements such as in \cite{bao2002inverse} or earthquake reconstruction such as in \cite{das1996inverse}. Our approach uses an SPDE method similar to \cite{lindgren2011explicit} to solve the non-point source problem in a Bayesian framework, providing the opportunity to explicitly incorporate linear regression models on the source function. This framework also allows us to incorporate covariance functions for the source, and to calculate uncertainties, in contrast to many existing inverse problems approaches, which often operate under either a prediction-only framework (e.g., \citealt{ayvaz2010linked}), or under more complex frameworks such as the ensemble Kalman filter (e.g., \citealt{stroud2010ensemble}). By incorporating our methodology and source-inference perspective into existing hierarchical frameworks \citep{wikle2010general}, process uncertainty and non-Gaussian observations could also be accommodated, although the use of a link-function would warp the mechanistic interpretations we seek to preserve in this study. Our method has a variety of possible applications, and we have shown that it is both accurate, and implementable using reproducible general-purpose FEM software \citep{alnaes2015fenics} that can easily accommodate arbitrary changes to the underlying PDE structure. 






\bibliographystyle{agsm}
\bibliography{spde}

\end{document}


\def\spacingset#1{\renewcommand{\baselinestretch}%
{#1}\small\normalsize} \spacingset{1}


\if1\blind
{
  \title{\bf Appendix to: Source Reconstruction for Spatio-Temporal Physical Statistical Models}
  \author{Connie Okasaki\thanks{
    This material is based upon work supported by the National Science Foundation Graduate Research Fellowship under Grant No. DGE-1762114}\hspace{.2cm}\\
    Quantitative Ecology and Resource Management Program, U of Washington,\\
    Mevin B. Hooten \\
    Department of Statistics and Data Sciences, University of Texas at Austin, \\
    and \\
    Andrew M. Berdahl\thanks{AMB was supported by the H. Mason Keeler Endowed Professorship in Sports Fisheries Management.} \\
    School of Aquatic and Fisheries Sciences, U of Washington}
  \maketitle
} \fi

\if0\blind
{
  \bigskip
  \bigskip
  \bigskip
  \begin{center}
    {\LARGE\bf Title}
\end{center}
  \medskip
} \fi

\noindent%
{\it Keywords:}  Inverse Problems, GMRF, Gaussian process, Basis function, SPDE

\pagebreak 



\section{Linear Regression on a Source Term}

A tool to improve inference on a source term, particularly given the relatively poor scale of spatial resolution implied by the slow observed convergence rate may be to model it using a linear regression. To do so we may assume
\begin{equation}
    f(\bm{{s}}) = \sum \beta_i x_i(\bm{{s}}) + \mathcal{E}(\bm{{s}}).
\end{equation}
Where $x_i(\bm{s})$ is assumed known, and $\mathcal{E}$ is some GMRF. Representing this equation within the FEM we obtain
\begin{equation}
    \bm{f} = \bm{X}\bm{\beta} + \bm{\eta}.
\end{equation}
If we place a Gaussian prior on $\bm{\beta}$ and assume it is independent of $\bm{\eta}$ then we obtain
\begin{equation}
\begin{split}
    \bm{\beta} & \sim N(\bm{\mu}_\beta,\bm{Q}_{\beta}^{-1}) \\
    \bm{\beta}\bm{X} & \sim N(\bm{X}\bm{\mu}_\beta,\bm{X}\bm{Q}_{\beta}^{-1}\bm{X}^T) \\
    \bm{\eta} & \sim N(\bm{\mu}_\eta,\bm{Q}_{\eta}^{-1}) \\
    \bm{f} & \sim N(\bm{X}\bm{\mu}_\beta + \bm{\mu}_\eta,\bm{X}\bm{Q}_{\beta}^{-1}\bm{X}^T + \bm{Q}_{\eta}^{-1}) \\
    & = N(\bm{X}\bm{\mu}_\beta + \bm{\mu}_\eta,(\bm{Q}_\eta - \bm{Q}_\eta\bm{X}(\bm{Q}_{\beta} + \bm{X}^T\bm{Q}_\eta\bm{X})^{-1}\bm{X}^T\bm{Q}_\eta)^{-1}).
\end{split}
\end{equation}
The matrix $\bm{X}$ is dense, but calculations involving the latter precision matrix can nevertheless be carried about by conducting a low-rank update to the Cholesky factor of $\bm{Q}_\eta$. Inference on $\bm{f}|\bm{y}$ can be carried out the same way as outlined in the main text, and inference on $\bm{\eta},\bm{\beta}|\bm{y}$ can be conducted similarly. We can write $\bm{y} = \bm{A}\bm{u} + \bm{\epsilon} = \bm{A}\bm{L}^{-1}\bm{K}(\bm{X}\bm{\beta}+\bm{\eta}) + \bm{\epsilon}$. For simplicity we will assume all prior means are 0. Then 
\begin{equation}
    \begin{bmatrix}\bm{\eta} \\ \bm{\beta} \\ \bm{y} \end{bmatrix} \sim N\left(\begin{bmatrix} 0 \\ 0 \\ 0\end{bmatrix},\begin{bmatrix}\bm{\Sigma}_\eta & 0 & \bm{\Sigma}_\eta\bm{K}^T\bm{L}^{-1}\bm{A}^T \\ 0 & \bm{Q}_{\beta} & \bm{Q}_{\beta}\bm{X}^T\bm{K}^T\bm{L}^{-1} \\ \bm{A}\bm{L}^{-1}\bm{K}\bm{\Sigma}_\eta  & \bm{A}\bm{L}^{-1}\bm{K}\bm{X}\bm{Q}_{\beta}  & \bm{A}\bm{\Sigma}_u\bm{A}^T +\sigma^2\bm{I}\end{bmatrix} \right)
\end{equation}
which, if we group $\bm{\theta} = \begin{bmatrix}\bm{\eta} \\ \bm{\beta}\end{bmatrix}$, then we may instead write this as
\begin{equation}
    \begin{bmatrix}\bm{\theta} \\ \bm{y} \end{bmatrix} \sim N\left(\begin{bmatrix}0 \\ 0\end{bmatrix},\begin{bmatrix}\bm{\Sigma}_\theta &  \bm{\Sigma}_\theta\bm{K}^T\bm{L}^{-1}\bm{A}^T \\ \bm{A}\bm{L}^{-1}\bm{K}\bm{\Sigma}_\theta & \bm{A}\bm{\Sigma}_u\bm{A}^T+\sigma^2\bm{I}\end{bmatrix} \right).
\end{equation}
Then we may write this in terms of its precision matrix
\begin{equation}
    \begin{bmatrix}\bm{\theta} \\ \bm{y} \end{bmatrix} \sim N\left(\begin{bmatrix}0 \\ 0\end{bmatrix},\begin{bmatrix}\bm{Q}_\theta+\frac{1}{\sigma^2}\bm{K}^T\bm{L}^{-1}\bm{A}^T\bm{A}\bm{L}^{-1}\bm{K} &  -\frac{1}{\sigma^2}\bm{K}^T\bm{L}^{-1}\bm{A}^T \\ -\frac{1}{\sigma^2}\bm{A}\bm{L}^{-1}\bm{K} & \frac{1}{\sigma^2}\bm{I}\end{bmatrix}^{-1} \right)
\end{equation}
Therefore we find that:
\begin{equation}
    \bm{\eta},\bm{\beta}|\bm{y} \sim N\left(\frac{1}{\sigma^2}\bm{Q}_{\eta,\beta|y}^{-1}\bm{K}^T\bm{L}^{-1}\bm{A}^T\bm{y},(\bm{Q}_u + \frac{1}{\sigma^2_\epsilon}\bm{K}^T\bm{L}^{-1}\bm{A}^T\bm{A}\bm{L}^{-1}\bm{K})^{-1}\right).
\end{equation}
An alternative way to conduct these calculations is to not skip straight to inferring $\bm{\eta}$, or even $\bm{f}$ but to first focus on inferring $\bm{u}$ and $\bm{\beta}$, then back-calculating $\bm{f}$ and $\bm{\eta}$ from these. In this case we obtain
\begin{align}
\begin{bmatrix}\bm{u} \\ \bm{\beta} \end{bmatrix}
& \sim \mathcal{N}\left(0,\begin{bmatrix}\bm{K}^{-1}\bm{L}\left(\bm{Q}^{-1} + \bm{XQ}_{\beta}^{-1}\bm{X}^T\right)\bm{LK}^{-T} & \bm{K}^{-1}\bm{LXQ}_{\beta}^{-1} \\ \bm{Q}_{\beta}^{-1}\bm{X^TLK}^{-T} & \bm{Q}_{\beta}^{-1}\end{bmatrix}\right) \nonumber \\
& \sim \mathcal{N}\left(0,\begin{bmatrix}\bm{K}^T\bm{L}^{-1}\bm{Q} \bm{L}^{-1}\bm{K} & -\bm{K}^T\bm{L}^{-1}\bm{Q}\bm{X} \\ -\bm{X}^T\bm{Q}\bm{L}^{-1}\bm{K} & \bm{Q}_\beta + \bm{X}^T\bm{Q}\bm{X} \end{bmatrix}^{-1}\right). \label{eq:linreg}
\end{align}
This has the advantage of allowing us to simplify our observation matrix and conduct inference using the conditional precision matrix $(\bm{Q}_{\bm{u},\bm{\beta}}+\frac{1}{\sigma^2}\bm{A}^T\bm{A})$. When inverting to obtain a Gibbs sampling step we reparameterize to $\bm{Q}^{-1}\sigma^2$ and $\bm{Q}_\beta^{-1}\sigma^2_\beta$:
\begin{align}
\begin{bmatrix}\bm{u} \\ \bm{\beta} \end{bmatrix}
& \sim \mathcal{N}\left(0,\begin{bmatrix}\bm{K}^{-1}\bm{L}\left(\sigma^2\bm{Q}^{-1} + \sigma^2_\beta\bm{XQ}_{\beta}^{-1}\bm{X}^T\right)\bm{LK}^{-T} & \sigma^2_\beta\bm{K}^{-1}\bm{LXQ}_{\beta}^{-1} \\ \bm{Q}_{\beta}^{-1}\bm{X^TLK}^{-T}/\sigma^2_\beta & \sigma^2_\beta\bm{Q}_{\beta}^{-1}\end{bmatrix}\right) \nonumber \\
& \sim \mathcal{N}\left(0,\sigma^2\begin{bmatrix}\bm{K}^{-1}\bm{L}\left(\bm{Q}^{-1} + v^2\bm{XQ}_{\beta}^{-1}\bm{X}^T\right)\bm{LK}^{-T} & v^2\bm{K}^{-1}\bm{LXQ}_{\beta}^{-1} \\ \bm{Q}_{\beta}^{-1}\bm{X^TLK}^{-T}/\sigma^2_\beta & v^2\bm{Q}_{\beta}^{-1}\end{bmatrix}\right) \nonumber \\
& \sim \mathcal{N}\left(0,\sigma^2\begin{bmatrix}\bm{K}^T\bm{L}^{-1}\bm{Q} \bm{L}^{-1}\bm{K} & -\bm{K}^T\bm{L}^{-1}\bm{Q}\bm{X} \\ -\bm{X}^T\bm{Q}\bm{L}^{-1}\bm{K} & \frac{1}{v^2}\bm{Q}_\beta + \bm{X}^T\bm{Q}\bm{X} \end{bmatrix}^{-1}\right). \label{eq:linreg}
\end{align}

\section{Accuracy Calculation}

Allowing in general for any matrices $\bm{{M}}(\bm{{Q}}_{\bm{{u}}} + \frac{\exp(\zeta)}{m\sigma^2_\epsilon}\bm{{A}}^T\bm{{A}})^{-1}\bm{{M}}$ to appear inside the trace, and differentiating with respect to $\zeta$, we obtain
\begin{equation}
\begin{split}
\dfrac{\partial\log L^2(\bm{{u}}|\bm{{y}})}{\partial \zeta} & \approx -\frac12\frac{\exp(\zeta)}{m\sigma^2}\frac{\rm{tr}(\bm{{I}}_{\rm int}\bm{{\Sigma}}(\nu)\bm{{A}}^T\bm{{A}}\bm{{\Sigma}}(\nu)\bm{{I}}_{\rm int})}{\rm{tr}(\bm{{I}}_{\rm int}\bm{{\Sigma}}(\nu)\bm{{I}}_{\rm int})} \\
\dfrac{\partial\log L^2(\bm{{f}}|\bm{{y}})}{\partial \zeta} & \approx -\frac12\frac{\exp(\zeta)}{m\sigma^2}\frac{\rm{tr}(\bm{{I}}_{\rm int}\bm{{L}}^{-1}\bm{{K}}\bm{{\Sigma}}(\nu)\bm{{A}}^T\bm{{A}}\bm{{\Sigma}}(\nu)\bm{{K}}^{T}\bm{{L}}^{-T}\bm{{I}}_{\rm int})}{\rm{tr}(\bm{{I}}_{\rm int}\bm{{L}}^{-1}\bm{{K}}\bm{{\Sigma}}(\nu)\bm{{K}}^{T}\bm{{L}}^{-T}\bm{{I}}_{\rm int})}. \\
\end{split}\label{eq:tr-formula}
\end{equation}

$\bm{{A}}^T\bm{{A}}$ (as well as $\bm{{I}}_{\rm int} = \bm{{I}}$) then we find that as $n\to \infty$, $(\bm{{Q}}_{\bm{{u}}}+\frac{n}{\sigma^2}\bm{{A}}^T\bm{{A}})^{-1} \approx \frac{\sigma^2}{n}(\bm{{A}}^T\bm{{A}})^{-1} = \frac{\sigma^2m}{\exp(\zeta)}(\bm{{A}}^T\bm{{A}})^{-1}$ such that 
\begin{equation}
\begin{split}
\dfrac{\partial\log L^2(\bm{{u}}|\bm{{y}})}{\partial \zeta} & \approx -\frac12\frac{\exp(\zeta)}{m\sigma^2}\frac{\rm{tr}(\bm{{\Sigma}}(\nu)\bm{{A}}^T\bm{{A}}\bm{{\Sigma}}(\nu))}{\rm{tr}(\bm{{\Sigma}}(\nu))} \\
& = -\frac12\frac{\exp(\zeta)}{m\sigma^2}\left(\left(\frac{\sigma^2m}{\exp(\zeta)}\right)^2\rm{tr}((\bm{{A}}^T\bm{{A}})^{-1})\right)\left(\frac{\sigma^2m}{\exp(\zeta)}\rm{tr}((\bm{{A}}^T\bm{{A}})^{-1})\right)^{-1}\\
& = -\frac12, \\
\dfrac{\partial\log L^2(\bm{{f}}|\bm{{y}})}{\partial \zeta} & \approx -\frac12. \\
\end{split}
\end{equation}

\section{Dirichlet Boundary Conditions}
Although Neumann boundary conditions on $\bm{u}$ are most common, Dirichlet boundary conditions on $\bm{u}$ can be incorporated into our approach, as can boundary conditions on $\bm{f}$. Using the basic FEM method, Neumann boundary conditions are considered ``natural,'' that is, the solution to $\bm{Ku} = \bm{Lf}$ will satisfy these BCs without any additional input. On the other hand Dirichlet boundary conditions are considered ``essential,'' meaning that we must impose these externally. Note that more sophisticated FEM methods may have different natural/essential BCs, so that these terms are not simply synonymous with Neumann/Dirichlet.

To demonstrate how to incorporate Dirichlet boundary conditions, suppose that our basic Neumann BC generates the FEM formula 
\begin{align}
\bm{K}\bm{u} = \bm{L}\bm{f}
\end{align}
Our Dirichlet boundary condition can be expressed by splitting $\bm{u}$ into component vectors $\bm{u}_{\partial\Omega}$ or $\bm{u}_1$ and $\bm{u}_{\Omega}$ or $\bm{u}_2$ representing the internal and external degrees of freedom (i.e., nodes/basis functions in our cases). This makes our FEM equation
\begin{align}
\begin{bmatrix} \bm{K}_{11} & \bm{K}_{12} \\ \bm{K}_{21} & \bm{K}_{22}\end{bmatrix}\begin{bmatrix} \bm{u}_1 \\ \bm{u}_2\end{bmatrix} = \begin{bmatrix}\bm{L}_{11} & \bm{L}_{12} \\ \bm{L}_{21} & \bm{L}_{22}\end{bmatrix}\begin{bmatrix}\bm{f}_1 \\ \bm{f}_2\end{bmatrix}.
\end{align}
To impose a Dirichlet boundary condition of the form $\bm{Bu}_{\partial\Omega} = \bm{c}$, we replace the first block-rows of this equation
\begin{align}
\begin{bmatrix} \bm{Q}_{\beta} & \bm{0} \\ \bm{K}_{21} & \bm{K}_{22}\end{bmatrix}\begin{bmatrix} \bm{u}_1 \\ \bm{u}_2\end{bmatrix} = \begin{bmatrix}\bm{I} & \bm{0} \\ \bm{L}_{21} & \bm{L}_{22}\end{bmatrix}\begin{bmatrix}\bm{c}
 \\ \bm{f}_2\end{bmatrix}.
\end{align}
The first set of equations determines $\bm{u}_1$, assuming that $\bm{Q}_{\beta}$ is invertible. Then we are left with
\begin{equation}
\bm{K}_{21}\bm{u}_1 + \bm{K}_{22}\bm{u}_2 = \bm{L}_{21}\bm{c} + \bm{L}_{22}\bm{f}_2.
\end{equation}
Assuming that $\bm{c}$ is a constant vector, we are left with a simple offset to the mean of $
\bm{u}_2$, i.e.,
\begin{align}
\bm{u}_2 & = \bm{\mu}+\bm{K}_{22}^{-1}\bm{L}_{22}\bm{f}_2, \\
\bm{\mu} & = \bm{L}_{21}\bm{c} - \bm{K}_{21}\bm{u}_1 \\
& = (\bm{L}_{21} + \bm{K}_{21}\bm{Q}_{\beta}^{-1})\bm{c},
\end{align}
In the event that $\bm{c} = \bm{0}$, this offset is also $\bm{0}$ and we are left with a lower-dimensional version of the same FEM equation. In the event that $\bm{c}$ is itself a lower-dimensional GMRF rather than a constant, then instead of a mean offset, this results in a more complex Gaussian random field with covariance 
\begin{equation}
(\bm{L}_{21} + \bm{K}_{21}\bm{Q}_{\beta}^{-1})\bm{\Sigma}_c(\bm{L}_{21} + \bm{K}_{21}\bm{Q}_{\beta}^{-1})^T + \bm{K}_{22}^{-1}\bm{L}_{22}\bm{\Sigma}_{f2}\bm{L}_{22}^T\bm{K}_{22}^{-T},
\end{equation}
This can be inverted using the Woodbury matrix formula, giving us
\begin{align}
\hspace{-1cm}\bm{Q}_{f1} & = \bm{Q}_{f2} - \bm{Q}_{f2}(\bm{L}_{21} + \bm{K}_{21}\bm{Q}_{\beta}^{-1})(\bm{Q}_c + (\bm{L}_{21} + \bm{K}_{21}\bm{Q}_{\beta}^{-1})^T\bm{Q}_{f2}(\bm{L}_{21} + \bm{K}_{21}\bm{Q}_{\beta}^{-1}))^{-1}
(\bm{L}_{21} + \bm{K}_{21}\bm{Q}_{\beta}^{-1})^T\bm{Q}_{f2}.
\end{align}
With the simplifying assumption that we replace $\bm{L}$ with $\tilde{\bm{L}}$, we arrive at $\bm{L}_{21} = \bm{0}$ so that we obtain
\begin{align}
\tilde{\bm{Q}}_{f2} & = \bm{Q}_{f2} - \bm{Q}_{f2}\bm{K}_{21}\bm{Q}_{\beta}^{-1}(\bm{Q}_c + \bm{K}_{21}^T\bm{Q}_{\beta}^{-T}\bm{Q}_{f2}\bm{Q}_{\beta}^{-1}\bm{K}_{21})^{-1}\bm{Q}_{\beta}^{-1}\bm{K}_{21}^T\bm{Q}_{f2}.
\end{align}

Assuming that $\bm{c}$ is much lower-dimensional than $\bm{f}_1$, this may be Cholesky decomposed by first decomposing $\bm{Q}_{f2}$ and then computing a low-rank update to this decomposition.

\section{Stabilization}

In the Gibbs step of our calculations we use a version of the Sherman-Morrison-Woodbury matrix formula to calculate the following using sparse matrix operations:
\[
d^T(\bm{O}\bm{\Sigma}\bm{O}^T + \sigma^2\bm{I})^{-1}d = d^T\left(\frac{1}{\sigma^2}\bm{I} - \frac{1}{\sigma^4}\bm{O}(\bm{Q} + \frac{1}{\sigma^2}\bm{O}^T\bm{O})^{-1}\bm{O}^T\right)d
\]
One way to stabilize this is by instead using the extended matrix:
\[
-\begin{bmatrix}d \\ 0\end{bmatrix}^T\begin{bmatrix}-\sigma^2\bm{I} & \bm{O} \\ \bm{O}^T & \bm{Q}\end{bmatrix}^{-1}\begin{bmatrix}d \\ 0\end{bmatrix}
\]
Using block matrix inversion, this produces an equivalent inner product. Then, if we can calculate a block triangular factor for this matrix we will be potentially able to conduct calculations in a stabler manner:
\[
\begin{bmatrix}-\sigma^2\bm{I} & \bm{O} \\ \bm{O}^T & \bm{Q}\end{bmatrix} = \begin{bmatrix}i \sigma I & 0 \\ -\frac{i}{\sigma}O^T & C\end{bmatrix}\begin{bmatrix}i \sigma I & -\frac{i}{\sigma}O \\ 0 & C^T\end{bmatrix}
\]